\documentclass[showpacs,preprintnumbers]{revtex4}
\usepackage{graphicx}
\usepackage{color}

\def\bc{\begin{center}}
\def\ec{\end{center}}

\def\beq{\begin{equation}}
\def\eeq{\end{equation}}

\begin{document}

\title{Tunable quantum entanglement of three qubits in a non-stationary cavity}
\author{Mirko Amico$^{1,2}$, Oleg L. Berman$^{1,2}$ and Roman Ya. Kezerashvili$^{1,2}$}
\affiliation{\mbox{$^{1}$Physics Department, New York City College
of Technology, The City University of New York,} \\
Brooklyn, NY 11201, USA \\
\mbox{$^{2}$The Graduate School and University Center, The
City University of New York,} \\
New York, NY 10016, USA}
\date{\today}

\date{\today}

\begin{abstract}

We investigate the tunable quantum entanglement and the
probabilities of excitations in a system
of three qubits in a non-stationary cavity due to the dynamical
Lamb effect, caused by non-adiabatic fast change of the boundary
conditions of the cavity. The transition amplitudes and the
probabilities of excitation of qubits  due to the dynamical Lamb
effect have been evaluated. The conditional concurrence and the
conditional residual tangle for each fixed amount of created photons
are introduced and calculated as measures
 of the pairwise or three-way dynamical quantum entanglement of the qubits. We also give a prescription on how to increase the values of those quantities by
 controlling the frequency of the cavity photons.
 A physical realization of the system with three superconducting qubits, coupled to a coplanar waveguide entangled due to the non-adiabatic fast change of boundary conditions of the cavity is proposed.

\end{abstract}


\maketitle

\section{Introduction}

\label{intro}

Since the early days of quantum theory, there has been a lot
controversy on its interpretation and its implications in our
understanding of what reality is. One of the most renown critics of
the theory was Einstein, who, together with his colleagues Podolski
and Rosen (EPR), wrote an article \cite{epr} which started a serious
discussion \cite{eprbohr} on the completeness of the theory. In
particular, they demonstrated through a thought experiment that
either: quantum mechanics is an incomplete theory or space-like
separated events can influence one another. A few decades later, in
an attempt to settle the argument raised by EPR, Bell showed
\cite{bell} that the predictions made in the framework of local
realistic theories were generally not in agreement with the ones
made in the framework of quantum mechanics. More recently,
Greenberger, Horne and Zeilinger (GHZ) proved in Refs. \cite{ghz2,
ghz} that for more complicated setups than the two-bodies
experiment, like a three-bodies experiment, one can never use local
realistic models to make definite predictions. Thus making quantum
mechanics the only framework where one is able to do that. In their
own words: ''\dots with the appropriate 4-particle (or even
3-particle) system, all one must do is prove that quantum theory
holds experimentally, and then we know that it cannot be classically
duplicated, \dots". This means that one needs only to verify the
predictions of quantum mechanics for such entangled many-body system
to prove the validity of quantum theory.

The latter motivated us to analyze a system of three qubits coupled to a superconducting line, playing the
role of a cavity, entangled due to the fast (non-adiabatic) change of
boundary conditions of the cavity. In this kind of setup there is the possibility
of tuning the quantum entanglement between the qubits by controlling the frequency of
the cavity photons. Being an entangled three-body system, based on
the results \cite{ghz2, ghz}, one only needs to verify
experimentally that quantum mechanics gives the correct predictions
to prove its validity. This will be another proof that quantum
mechanics has to be used to describe our universe and no local
realistic theory will actually work. The quantum entanglement of the
qubits arises due to the dynamical Lamb effect (DLE), first
described in Ref. \cite{Lozovik}, which is the photon-less
(parametric) excitation of the qubits in the cavity through the
modulation of their Lamb shift achieved by non-adiabatic change of
the boundary conditions. In this quantum phenomenon, virtual
qubit-photon states become real due to the change in boundary
conditions. The other interesting quantum effect is the dynamical
Casimir effect (DCE), which is the creation of real cavity photons
from vacuum fluctuations due to the time mismatch between vacuum
modes of the photon field of the cavity, when the cavity length is
changed non-adiabatically \cite{Moore}. In this paper, an experiment
based on the excitation of qubits due to the DLE caused by
non-adiabatic change of boundary conditions of the cavity is
proposed. It is important to note that the quantum entanglement
which arises in this way is entirely due to the non-adiabatic change
of boundary conditions and does not involve any exchange of photons
between the qubits.

All calculations are made with a specific experimental setup in mind. We consider superconducting qubits coupled
to a coplanar waveguide whose length can be tuned quickly enough to satisfy the condition for non-adiabatic change of boundaries of the cavity.
As explained in details in Ref. \cite{Nori_Nature}, a superconducting qubit can easily be made from a superconducting island connected to a reservoir
of Cooper pairs (in the bulk) through a Josephson junction. While for the cavity, a one-dimensional transmission line resonator made from
a superconducting coplanar waveguide can be used. This system works as a model for atoms in an optical cavity. More details about the use of
 superconducting circuits to carry out experiments in cavity quantum electrodynamics can be found in Refs. \cite{Girvin,Lozovik_2015, Girvin_Nature}.
 The setup considered allows for an easier tuning of the cavity and qubit's parameters, for instance, the cavity length
  can be modulated fast enough to have non-adiabatic changes of boundary conditions.
  So, it seems to be an optimal apparatus to explore new phenomena that can arise in cavity quantum electrodynamics
  when a greater control over the parameters of the experiment is achieved. Indeed, the setup enables to reproduce conditions that would hardly be physically realizable with real atoms and mirrors, like the non-adiabatic change
  in the boundary conditions which gives rise to new interesting quantum effects such as the dynamical Casimir effect and the dynamical Lamb effect.
  The DCE, for example, was observed for the first time in a system of superconducting qubits in Ref. \cite{Wilson}.

The quantum entanglement of the system of three qubits is studied by calculating the conditional residual tangle and the conditional concurrence. The conditional residual tangle is a new quantity introduced in this paper: it is the residual tangle, defined in Ref. \cite{Coffman} as a measure of the quantum entanglement between three qubits, for a certain fixed number of created photons in the cavity.
The conditional concurrence, which was previously introduced in Ref. \cite{berman}, measures the pairwise quantum entanglement between the qubits for a fixed number of created photons in the cavity.
If the three qubits are three-way entangled, for example,
if they are in a GHZ state $\frac{1}{\sqrt{2}}\left(\left|000\right\rangle\pm\left|111\right\rangle\right)$, a measure on the state
of one of them determines the state of all the others. But a system of three qubits could also show pairwise entanglement, as in
a W state $\frac{1}{\sqrt{3}}\left(\left|100\right\rangle\pm\left|010\right\rangle\pm\left|001\right\rangle\right)$, so that a measurement on
 one of the qubit leaves the other ones still entangled. These two types of entanglement are complementary, so both kind of entanglement cannot
  be dominant at the same time and the system will mainly show either three-way entanglement or two-way entanglement.
We are mostly interested in the case where all three qubits are entangled, characterized by having a residual tangle different from zero,
in order to find out more about the state of the system which would allow us to make predictions that can be used to experimentally prove the validity of quantum theory.
Also, the amplitudes of transition and the probabilities of excitation of the qubits due to the non-adiabatic change of boundary conditions of the cavity are calculated.
These quantities are necessary to determine the conditional concurrence and the conditional residual tangle of the qubits.
The results obtained show that it is possible to tune the values
of the quantities of interest by changing the frequency of the photons in the final state.
For instance, the quantum entanglement between the qubits could be increased in this way.
Although, a note of caution must be made: all results were found in the framework of time-independent
perturbation theory so the parameters cannot be changed arbitrarily but the requirements of applicability of this approximation must still be satisfied.

The paper is organized in the following way: in Sec.~\ref{ham} we put down the basis for the analysis of the case of interest. Once the Hamiltonian of the system is established, the amplitudes of transition and the probabilities of excitation of qubits are calculated. In Sec. \ref{entan}, we calculate the quantities which give a measure of the entanglement between the qubits. The expressions of the conditional residual tangle and the conditional concurrence for each number of created photons in the cavity are found and an estimate of those quantities for a particular physical realization is given.
The main results and their implications are summarized in Sec.~\ref{conclusions}.

\section{The Hamiltonian of three qubits coupled to a cavity}

\label{ham}

In this section, the Hamiltonian for the system of three qubits in a non-stationary cavity is presented, the transition amplitudes and the probabilities of excitation of the qubits for the system of three qubits and $n$ photons due to non-adiabatic change of the boundary conditions of the cavity are also calculated.

One can write the Hamiltonian of three qubits coupled to an optical cavity with changing boundary conditions following Ref.~\cite{berman}, where the two qubits case is considered. The generalization for the case of a three qubits system reads ($\hbar = 1$):
\begin{eqnarray}
\label{ham22} \hat{\mathcal{H}} = \hat{H}  +  i \frac{\dot{\omega}(t)}{4 \omega(t)}\left(a^{2} - a^{\dagger 2}\right),
\end{eqnarray}
$\omega(t)$ is the frequency of the electromagnetic mode in the cavity, $a^{\dagger }$ and $a$ are
creation and annihilation operators for cavity photons, and $\hat{H}$ is the Hamiltonian in the case of a stationary cavity
\begin{eqnarray}
\label{ham223}
\hat{H} = \hat{H}_{0} + \hat{V}_{total}.
\end{eqnarray}
In Eq. (\ref{ham22}), $\hat{H}_{0}$ is the Hamiltonian of three non-interacting qubits and $\hat{V}_{total}$ is the qubit-photon interaction. They are:
\begin{eqnarray}
\label{ham022} \hat{H}_{0} =
E_{0}\sum_{j=1}^{3}\frac{1+\hat{\sigma}_{3j}}{2} + \omega a^{\dagger}a,
\end{eqnarray}
\begin{eqnarray}
\label{V22} \hat{V}_{total} = \lambda \sum_{j=1}^{3}
\left(\hat{\sigma}_{j}^{+} + \hat{\sigma}_{j}^{-}\right)\left(a +
a^{\dagger}\right),
\end{eqnarray}
where $E_{0}$ is the qubit transition frequency and $\lambda $ is the
strength of the artificial qubit-photon coupling, while $\hat{\sigma}^{+}=\left(
\hat{\sigma}_{1}+i\hat{\sigma}_{2}\right) /2$ and $\hat{\sigma}%
^{-}=\left( \hat{\sigma}_{1}-i\hat{\sigma}_{2}\right) /2$, are defined via the Pauli matrices $\hat{\sigma}%
_{1}$, $\hat{\sigma}_{2}$, $\hat{\sigma}_{3}$.

From Eq.~(\ref{V22}) we see that $\hat{V}_{total}$ can be split into two
parts: $\hat{V}_{RWA} =  \lambda \sum_{j=1}^{3}
\left(\hat{\sigma}_{j}^{+}a + \hat{\sigma}_{j}^{-}a^{\dagger}\right) \ $that includes the
qubit-photon interaction in the rotating wave approximation (RWA) which does
not change the number of the excitations in the system and $\hat{V}=\lambda
\sum_{j=1}^{3}\left( \hat{\sigma}_{j}^{+}a^{\dagger }+\hat{\sigma}%
_{j}^{-}a\right) $ that includes the terms beyond the RWA, which can
increase or decrease the number of the excitations in the system:%
\begin{eqnarray}
\hat{V}_{total}=\hat{V}_{RWA}+\hat{V}.
\end{eqnarray}

Since the DLE is related to the parametric excitation of the qubit and the
creation of cavity photons \cite{Lozovik}, only $\hat{V}$ influences the
DLE. Indeed, the other term does not introduce new excitations into the
system but preserves the excitation number of qubits and photons. Therefore
in the framework of time-independent perturbation theory one is taking $%
\hat{V}$ as the perturbation to the ground state Hamiltonian $\hat{H}_{0}$.


Now let's calculate the transition amplitudes $A_{n;m}^{(L)}$ and the probabilities of excitation of the qubits $w_{n;m}^{(L)}$. In this notation $n$ counts the number of excited photons, $m$ counts the number of excited qubit and the superscript $(L)$ specifies that only the terms which contribute to the dynamical Lamb effect are taken into account.
It is clear that all probabilities that involve the same qubit excitation number are going to be the same. This is due to the fact that the qubits are indistinguishable, so when one -or more- qubit is excited nobody can tell which one of the three really is. Below, the case which gives rise to the dynamical Lamb effect is considered: the system in its ground state is excited through a non-adiabatic change of boundary conditions. As explained earlier, this change in boundary conditions can excite the qubits and create new cavity photons. This is taken into account by explicitly considering two different frequency of the cavity photons between initial and final states. In fact, the change in cavity length reflects in a change of the electromagnetic modes allowed in the cavity, thus causing a frequency shift of the created photons.

In general, the amplitude $A_{n;1}^{(L)}$ for
excitation of one qubit and creation of  $n$ cavity photons is
given by
\begin{eqnarray}
\label{An1} A_{n;1}^{(L)} =  _{\lambda \omega_{2}}\left\langle
n;100 \right| \left. 0;000 \right\rangle_{\lambda \omega_{1}}=  _{\lambda \omega_{2}}\left\langle
n;010 \right| \left. 0;000 \right\rangle_{\lambda \omega_{1}}=  _{\lambda \omega_{2}}\left\langle
n;001 \right| \left. 0;000 \right\rangle_{\lambda \omega_{1}}.
\end{eqnarray}

\begin{eqnarray}
A_{1;1}^{(L)} = \frac{1}{3\lambda}\left(- E_{L,0}^{(\lambda)}(\omega_{2}) + E_{L,0}^{(\lambda)}(\omega_{1})\right),
\end{eqnarray}
where the expressions for the $E_{L,0}^{(\lambda)}(\omega) $ are give by Eqs. (\ref{Lamb}) in Appendix \ref{appendixa}.
The latter result means that a qubit is excited only when a cavity photon is created.

The amplitude $A_{n;2}^{(L)}$, for excitation of two qubits and
creation of  $n$ cavity photons is given by
\begin{eqnarray}
\label{An2} A_{n;2}^{(L)} =  _{\lambda \omega_{2}}\left\langle
n;110 \right| \left. 0;000 \right\rangle_{\lambda \omega_{1}} =  _{\lambda \omega_{2}}\left\langle
n;101 \right| \left. 0;000 \right\rangle_{\lambda \omega_{1}}=  _{\lambda \omega_{2}}\left\langle
n;011 \right| \left. 0;000 \right\rangle_{\lambda \omega_{1}}.
\end{eqnarray}
The latter is different from zero only for $n=0,2$ created photons:
\begin{eqnarray}
A_{0;2}^{(L)} = \frac{2}{9\lambda^2}E_{L,3}^{(\lambda)}(\omega_{2})  E_{L,0}^{(\lambda)}(\omega_{1}), \\
\label{A22} A_{2;2}^{(L)} = -\frac{2\sqrt{2}}{9\lambda^2}E_{L,0}^{(\lambda)}(\omega_{2})  E_{L,0}^{(\lambda)}(\omega_{1}).
\end{eqnarray}

For the excitation of two qubits there are two possibilities: no cavity photons are created in the process or two cavity photons are created. It can be noted that the transition amplitude $A_{0;2}^{(L)}$, which does not involve the creation of cavity photons, increases when the frequency of the photons $\omega_2$  in the final state approaches the qubit's excitation energy $E_0$.

Finally, the amplitude $A_{n;3}^{(L)}$ for excitation of all three qubits and
creation of  $n$ cavity photons is given by
\begin{eqnarray}
\label{An3} A_{n;3}^{(L)} =  _{\lambda \omega_{2}}\left\langle
n;111 \right| \left. 0;000 \right\rangle_{\lambda \omega_{1}}.
\end{eqnarray}

This last result, $A_{n;3}^{(L)} = 0$, shows that in this framework it is impossible for the system to undergo such transition. It can be taken as a hint that the probability of this event will be noticeably smaller than the ones with amplitude different from zero.

Now, let's turn to the calculation of the probabilities of DLE using the results just found. We are interested in the total probability of excitation from the ground state to one of the qubits' excited states. So, first, the probabilities of excitation of the qubits with respect to a fixed number of created photons are calculated and then the sum over all possible number of photons is taken.

For the case of excitation of one qubit the probability $w_{n;1}^{(L)}$ is
\begin{eqnarray}
\label{wn1}  w_{n;1}^{(L)} = \sum_{n=0}^{\infty} \left|A_{n;1}^{(L)}\right|^{2} = \left|A_{1;1}^{(L)}\right|^{2} =
\frac{1}{9\lambda^2}\left(- E_{L,0}^{(\lambda)}(\omega_{2}) + E_{L,0}^{(\lambda)}(\omega_{1})\right)^2.
\end{eqnarray}

Substituting the values of the parameters that appear in Eq. (\ref{wn1}) from \cite{Delsing} and \cite{Johansson} it is possible to give an estimate of the probability of excitation of one qubit due to the dynamical Lamb effect. In the calculations below we will use: $\omega_{1} = 2\pi \times 5 \ \mathrm{GHz}$,
$E_{0} = 2\pi \times 3.721 \  \mathrm{GHz}$, $\lambda = 0.04 \omega_{1} =  2\pi \times 0.2 \  \mathrm{GHz}$, $\omega_{2} = \omega_{1} - 0.25 \omega_{1}= 2\pi \times 3.75 \  \mathrm{GHz}$.

With this choice of parameters, the probability of excitation of one qubit due to the DLE is $ w_{n;1}^{(L)} = 1.47 \times 10^{-5} $. Since there aren't any possibly singular terms in $w_{n;1}^{(L)}$, it's not easy to increase its value by tuning the parameters of the setup.

The contribution of DLE to the probability of excitation of two qubits $w_{n;2}^{(L)}$ is given by
\begin{eqnarray}
\label{wn2}  w_{n;2}^{(L)} =  \frac{4}{81\lambda^4}E_{L,0}^{(\lambda)}(\omega_{1})\left(E_{L,3}^{(\lambda)}(\omega_{2})^{2} + 2E_{L,0}^{(\lambda)}(\omega_{2})^{2}\right).
\end{eqnarray}

According to Eqs.~(\ref{Lamb}), $ w_{n;2}^{(L)}$ has a singularity when $\omega_2=E_{0}$. Therefore there is the possibility of choosing the value of the resonant frequency $\omega_2$ that yields the desired probability of excitation of two qubits. However, the frequency $\omega_2$ should be tuned carefully, so that perturbation theory is not broken and the predictions made in this framework are still trustworthy.

For the value of the parameters chosen earlier, one obtains a probability of excitation of two qubits $ w_{n;2}^{(L)}  = 0.1 $ .

Finally, we note that the probability of excitation of all three quits from the ground state $w_{n;3}^{(L)}=0$ because the transition is not allowed within our framework. Since the probability of excitation of two qubits due to the dynamical Lamb effect in Eq. (\ref{wn2}) is not equal to the product of the probabilities of excitation of a single qubit from Eq. (\ref{wn1}), the quantum entanglement for two qubits in the system of three qubits occurs due to the dynamical Lamb effect. The summary of transition amplitudes and probabilities for all possible final state are listed in Table \ref{amp}.

\begin{table}[ht]
\caption{Summary of transition amplitudes and probabilities of excitation for all possible final states} 
\centering 
\begin{tabular}{c c c } 
\hline\hline 
\\
 Final state \, \, \,  & Transition amplitude $A_{n;m}^{(L)}$ \,\,\, \, \, \, \,   & Probability of excitation $w_{n;m}^{(L)}$ \\  \hline \\
 $ \left|n;0 \right\rangle$ & $ A_{2;0}^{(L)} = -\frac{3\sqrt{2}\lambda^{2}}{\left(\omega_{1} + E_{0} \right)\left(\omega_{2} - E_{0} \right)}$ & $w_{n;0}^{(L)} = \left(\frac{3\sqrt{2}\lambda^{2}}{\left(\omega_{1} + E_{0} \right)\left(\omega_{2} - E_{0} \right)}\right)^2  $  \\  [1ex]
\\
$ \left|n;1 \right\rangle$ & $ A_{1;1}^{(L)} = \lambda\left(\frac{1}{\omega_{2} + E_{0}} - \frac{1}{\omega_{1} + E_{0}}\right)$ & $ w_{n;1}^{(L)} = \lambda^{2}\left(\frac{1}{\omega_{2} + E_{0}} - \frac{1}{\omega_{1}
+ E_{0}}\right)^{2} $   \\  [1ex]
\\
 $ \left|n;2\right\rangle$ & $A_{0;2}^{(L)} = \frac{2\lambda^2}{\left(\omega_{2} - E_{0}\right)\left(\omega_{1} + E_{0}\right)}$  & \\
\\
 &  $ A_{2;2}^{(L)} = - \frac{2\sqrt{2}\lambda^2}{\left(\omega_{2} + E_{0}\right)\left(\omega_{1} + E_{0}\right)} $ & $ \, \,  w_{n;2}^{(L)} = \left(\frac{2\lambda^2}{\left(\omega_{2} - E_{0}\right)\left(\omega_{1} + E_{0}\right)}\right)^{2} + \left(\frac{-2\sqrt{2}\lambda^2}{\left(\omega_{2} + E_{0}\right)\left(\omega_{1} + E_{0}\right)}\right)^{2} $ \\  [1ex]
\\
$ \left|n;3\right\rangle$ & $  A_{n;3}^{(L)} = 0 $ & $  w_{n;3}^{(L)} =0 $  \\ [1ex] 
\\
\hline \hline 
\end{tabular}
\label{amp} 
\end{table}

\section{Quantum entanglement for three qubits}

In this section we study the entanglement between qubits which arises when we change the boundary conditions non-adiabatically. When the system is excited from the ground state to an excited state, the qubits become entangled due to the non-adiabatic change of boundary conditions of the cavity. Different type of entanglement between the qubits in the cavity may arise. For three qubits there are two possibilities: a pairwise entanglement between two of the three qubits or a three-way entanglement between the three qubits altogether. As we shall see, only one type of entanglement can be prevalent for a given final state of the system.

\label{entan}

\subsection{Conditional residual tangle}

Let's consider the first case where all three qubits are entangled. A measure of the quantum entanglement between three qubits was defined in~\cite{Coffman}, and it is called residual tangle $\tau_{ABC}$, where $A, B$ and $C$ are labels for the three qubits. Although, as discussed earlier, the qubits are actually undistinguishable and so the labeling does not really matter, we found that naming the different qubits makes some of the upcoming formulas clearer.

Let's introduce here a new quantity, the conditional residual tangle, which is strictly related to the residual tangle. The conditional residual tangle is the residual tangle of the system for a fixed number of created photons in the cavity. The value of this quantity for all possible final states of the qubits for each number of created photons will then be calculated.

In general, a three qubits state can be written as

\begin{eqnarray}
\label{Phi} \left|\Phi\right\rangle = \sum_{ijk} a_{ijk}\left|ijk\right\rangle, \; \; \; \ \, \, \, \,\,\,\, \, i,j,k=0,1,
\end{eqnarray}
where

\begin{eqnarray}
a_{000} = A_{n;0}^{(L)}, \ \ \  a_{100} = A_{n;1}^{(L)},  \ \ \   a_{010} = A_{n;1}^{(L)},  \ \ \   a_{001} = A_{n;1}^{(L)}, \nonumber \\ a_{110} = A_{n;2}^{(L)},  \ \ \   a_{101} = A_{n;2}^{(L)},  \ \ \   a_{011} = A_{n;2}^{(L)},  \ \ \   a_{111} = A_{n;3}^{(L)}.
\end{eqnarray}

Following Ref. \cite{Coffman}, the conditional residual tangle for a fixed number of created photons $n$ can be written as:

\begin{eqnarray}
\tau_{ABC}^{\left|n\right\rangle} = 4\left|d_{1}-2d_{2}-4d_{3}\right|,
\end{eqnarray}
where

\begin{eqnarray}
d_{1} = a_{000}^{2}a_{111}^{2} + a_{001}^{2}a_{110}^{2} + a_{010}^{2}a_{101}^{2} + a_{100}^{2}a_{011}^{2}, \nonumber
\end{eqnarray}
\begin{eqnarray}
d_{2} = a_{000}a_{111}a_{011}a_{100} + a_{000}a_{111}a_{101}a_{010} + a_{000}a_{111}a_{110}a_{001} + \nonumber \\ + a_{001}a_{100}a_{101}a_{010} + a_{011}a_{100}a_{110}a_{001} + a_{101}a_{010}a_{110}a_{001}, \nonumber
\end{eqnarray}
\begin{eqnarray}
d_{3} =  a_{000}a_{110}a_{101}a_{011} + a_{111}a_{001}a_{010}a_{100}.
\end{eqnarray}

Proceeding with the calculation of the conditional residual tangle for each number of created photons, the following results were found. For zero created photons, $n=0$, one finds that there is no residual tangle between the qubits:

\begin{eqnarray}
\tau_{ABC}^{\left|0\right\rangle} =0.
\end{eqnarray}

Thus, the three qubits are not all simultaneously entangled and may only be pairwise entangled with one another.

For one created photon, $n=1$:

\begin{eqnarray}
\tau_{ABC}^{\left|1\right\rangle} =0.
\end{eqnarray}

Therefore, also the final state associated with the creation of a cavity photon shows no residual tangle.

For two created photons, $n=2$:

\begin{eqnarray}
\label{tau2}
\tau_{ABC}^{\left|2\right\rangle} = 16\left|d_{3}\right| = 16\left|A_{2;0}^{(L)}A_{2;2}^{(L)}A_{2;2}^{(L)}A_{2;2}^{(L)}\right|.
\end{eqnarray}

Substituting the expressions for the amplitude of creation of two photons with two qubit excitations $A_{2;2}^{(L)}$, and the amplitude of creation of two photons $A_{2;0}^{(L)}$ from Table \ref{amp} into Eq.~(\ref{tau2}), one gets the following result for the conditional residual tangle:

\begin{eqnarray}
\tau_{ABC}^{\left|2\right\rangle} = 16 \left(\frac{3\sqrt{2}\lambda^{2}}{\left(\omega_{1} + E_{0} \right)\left|\omega_{2} - E_{0} \right|}\right)\left(\frac{2\sqrt{2}\lambda^2}{\left(\omega_{2} + E_{0}\right)\left(\omega_{1} + E_{0}\right)}\right)^{3}.
\end{eqnarray}

We note that the latter expression increases when $\omega_{2}$ approaches $E_{0}$, that is when the frequency of the created photons in the cavity is close to the transition frequency of the qubits. Consequently, we have the opportunity to increase the value of the residual tangle by precisely tuning $\omega_2$. Using the same value for the parameters of the setup as done before, one gets: $\tau_{ABC}^{\left|2\right\rangle} = 5.62 \times 10^{-8}$.

Thus, the final state which shows entanglement between all three qubits is

\begin{eqnarray}
\label{3state}
\left|\bar\Phi\right\rangle = a_{000}\left|2;000\right\rangle  + a_{110}\left|2;110\right\rangle + a_{101}\left|2;101\right\rangle + a_{011}\left|2;011\right\rangle.
\end{eqnarray}

This result tells us that it is possible to excite the system so that it shows entanglement between three bodies. Thus, once the system is brought to this state, it is possible to carry out measurements involving three entangled bodies which, as shown by GHZ in Ref. \cite{ghz}, are crucial to test the validity of quantum theory.

The other values of the conditional residual tangle are found to be zero for any number of created photons greater than two. This is so, because the transition amplitudes for $n>2$ are all zero making all factors in the calculation of the conditional residual tangle vanish. The summary of calculations for the conditional residual tangle for each fixed $n$ of created photons in the cavity is given in Table \ref{tan}.

\begin{table}[ht]
\caption{Summary of the conditional residual tangle for each fixed number of created photons in the cavity $n$} 
\centering 
\begin{tabular}{c c} 
\hline\hline 
  Number of photons $n$ &  $\tau_{ABC}^{\left|n\right\rangle}$  \\  \hline
\\
 $ n=0 $ & $ 0 $  \\  [1ex]
\\
 $ n=1$ & $  0 $   \\  [1ex]
\\
 $ n=2$ & $ 16 \left(\frac{3\sqrt{2}\lambda^{2}}{\left(\omega_{1} + E_{0} \right)\left|\omega_{2} - E_{0} \right|}\right)\left(\frac{2\sqrt{2}\lambda^2}{\left(\omega_{2} + E_{0}\right)\left(\omega_{1} + E_{0}\right)}\right)^{3} $\\  [1ex]
\\
$ n>2 $ & $  0 $  \\ [1ex] 
\hline \hline 
\\
\end{tabular}
\label{tan} 
\end{table}

\subsection{Conditional concurrence}
Let's consider the second case where the qubits show pairwise entanglement between each other. Since the different type of entanglement are complementary, when one kind of entanglement is large the other one must be small. Accordingly, if the system shows a high degree of three-way entanglement it will have a low degree of pairwise entanglement and viceversa. This is easily seen from the formula relating these different types of entanglement found in~\cite{Coffman}, which is reported here:

\begin{eqnarray}
\label{23tang} \tau_{A(BC)}^{\left|n\right\rangle} = \tau_{AB}^{\left|n\right\rangle} + \tau_{AC}^{\left|n\right\rangle} + \tau_{ABC}^{\left|n\right\rangle},
\end{eqnarray}

where $\tau_{A(BC)}^{\left|n\right\rangle}$ is the conditional tangle between the qubit A and the entangled couple BC, $\tau_{AB}^{\left|n\right\rangle}$ and $\tau_{AC}^{\left|n\right\rangle}$ are the conditional concurrences for the qubits pair AB and AC and $\tau_{ABC}^{\left|n\right\rangle}$ is the conditional residual tangle of the three qubits ABC. Since the labeling doesn't matter, the formula is true for any permutation of the indices.

From now on, the conditional concurrence for $n$ created photons will be denoted as $C^{\left|n\right\rangle}$.
Similarly to what was done for the case of two qubits in Ref.~\cite{berman}, we calculate the conditional concurrence for each number of created photons. In order to do so, first the number of created photons $n$ is fixed, then a pair of qubits for which we calculate the conditional concurrence is chosen. The state of the third qubit is fixed (either to 1 or 0) when doing the calculation. If the three qubits were different from one another we would have to distinguish which qubit is entangled and which is not. In the case treated here, were the qubits are identical, we are not affected by this distinction. Thus we consider only one pair of qubits and all permutations of the indices will give the same results for the other possible pairs.

The following states are allowed:

\begin{eqnarray}
\label{Phi2} \left|\Phi_{AB0}\right\rangle = a\left|000\right\rangle + b\left|100\right\rangle + c\left|010\right\rangle + d\left|110\right\rangle, \nonumber \\ \nonumber \\
\left|\Phi_{AB1}\right\rangle = a'\left|001\right\rangle + b'\left|101\right\rangle + c'\left|011\right\rangle + d'\left|111\right\rangle,
\end{eqnarray}

where

\begin{eqnarray}
a = A_{n;0}^{(L)}, \ \ \  b = A_{n;1}^{(L)},  \ \ \   c = A_{n;1}^{(L)},  \ \ \   d = A_{n;2}^{(L)}, \nonumber \\ a' = A_{n;1}^{(L)},  \ \ \   b' = A_{n;2}^{(L)},  \ \ \   c' = A_{n;2}^{(L)},  \ \ \   d' = A_{n;3}^{(L)}.
\end{eqnarray}

The corresponding concurrences $C (\Phi_{AB0})$ and $C (\Phi_{AB1})$ defined in~\cite{Wootters_1997,Wootters_1998,Wootters} of the two possible states $\Phi$ are:

\begin{eqnarray}
\label{CPhi} C (\Phi_{AB0}) = 2 \left|ad - bc \right|,
\end{eqnarray}

\begin{eqnarray}
 C (\Phi_{AB1}) = 2 \left|a'd' - b'c' \right|.
\end{eqnarray}

Let's now calculate the conditional concurrence for each number of created photons $n$ for the different possible qubit pairs.
In the case where there are no created photons in the cavity, the conditional concurrence is

\begin{eqnarray}
C^{\left|0\right\rangle}_{AB0} = 0.
\end{eqnarray}

Thus, there is no entanglement between any pair of qubits if the third qubit is in the ground state and no cavity photons are created. The other possibility is that the third qubit is excited, the conditional concurrence is then

\begin{eqnarray}
C^{\left|0\right\rangle}_{AB1} = 2\left|-bc\right|= 2 \left|A_{0;2}^{(L)}A_{0;2}^{(L)}\right| =  2\left(\frac{2\lambda^2}{\left(\omega_{2} - E_{0}\right)\left(\omega_{1} + E_{0}\right)}\right)^{2}.
\end{eqnarray}

Therefore, if the third qubit is in the excited state and no cavity photons are created, there can be entanglement between the remaining pair of qubits.
The corresponding estimated conditional concurrence is $C^{\left|0\right\rangle}_{AB1} = 0.2 $. This value can be increased by tuning $\omega_2$ as the expression increases when $\omega_2 \rightarrow E_0$.

For the case of $n=1$ created photons, one gets

\begin{eqnarray}
C^{\left|1\right\rangle}_{AB0} = 2\lambda^2\left(\frac{1}{\omega_{2} + E_{0}} - \frac{1}{\omega_{1} + E_{0}}\right)^2.
\end{eqnarray}

Thus, when a photon is created in the cavity there can be entanglement between the pair of qubits, given the third qubit is in its ground state. An estimate for the conditional concurrence in this case gives: $C^{\left|1\right\rangle}_{AB0} = 2.95 \times 10^{-5}$.

When the third qubit is excited, one gets

\begin{eqnarray}
C^{\left|1\right\rangle}_{AB1} = 0.
\end{eqnarray}

Therefore, no entanglement is shared among the qubits.

For the conditional concurrence in the case of $n=2$ created photons we find

\begin{eqnarray}
C^{\left|2\right\rangle}_{AB0} = \frac{24\lambda^4}{\left|\omega_{2}^2 - E_{0}^2\right|\left(\omega_{1} + E_{0}\right)^2},
\end{eqnarray}

$C^{\left|2\right\rangle}_{AB0} = 2.33 \times 10^{-3}$. Also in this case, the value of the conditional concurrence can be increased by changing $\omega_2$.

When the third qubit is excited, one gets

\begin{eqnarray}
C^{\left|2\right\rangle}_{AB1} =  \frac{8\lambda^4}{\left(\omega_{2} + E_{0}\right)^2\left(\omega_{1} + E_{0}\right)^2},
\end{eqnarray}

and the corresponding numerical value is $C^{\left|2\right\rangle}_{AB1} = 3.02 \times 10^{-6}$.

The conditional concurrence is different from zero in both cases. Thus, any state associated with the creation of two photons in the cavity will always show some degree of pairwise entanglement of the qubits.

For $n>2$ created photons in the cavity, the conditional concurrence is always zero. That is due to the transition amplitudes in Eqs. (\ref{An1}), (\ref{An2}) and (\ref{An3}) vanishing when more than two photons are created.  The summary of the conditional concurrences for each fixed number $n$ of created photons in the cavity are given in Table \ref{conc}.

\begin{table}[ht]
\caption{Summary of the conditional concurrences for each fixed number of created photons in the cavity $n$} 
\centering 
\begin{tabular}{c c c } 
\hline\hline 
 Number of photons $n$ &  $C^{\left|n\right\rangle}_{AB0}$ &  $C^{\left|n\right\rangle}_{AB1}$ \\  \hline
\\
 $ n=0 $ & $ 0 $ & $ 2\left(\frac{2\lambda^2}{\left(\omega_{2} - E_{0}\right)\left(\omega_{1} + E_{0}\right)}\right)^{2}  $  \\  [1ex]
\\
 $ n=1$ & $ 2\lambda^2\left(\frac{1}{\omega_{2} + E_{0}} - \frac{1}{\omega_{1} + E_{0}}\right)^2$ & $ 0 $   \\  [1ex]
\\
 $ n=2$ & $\frac{24\lambda^4}{\left|\omega_{2}^2 - E_{0}^2\right|\left(\omega_{1} + E_{0}\right)^2} $ & $ \frac{8\lambda^4}{\left(\omega_{2} + E_{0}\right)^2\left(\omega_{1} + E_{0}\right)^2} $ \\  [1ex]
\\
$ n>2 $ & $  0 $ & $  0 $  \\ [1ex] 
\hline \hline 
\end{tabular}
\label{conc} 
\end{table}

\section{Discussion and Conclusions}

\label{conclusions}

We have investigated a system of three qubits coupled to a cavity where quantum entanglement between the qubits arises due to the non-adiabatic change of boundary conditions of the cavity itself. The parametric (photonless) excitation of qubits through the modulation of their Lamb shift caused by the non-adiabatic change of boundary conditions is a new, yet to be observed, quantum phenomena called dynamical Lamb effect. By exciting the qubits in this way, the quantum entanglement between them does not arise due to the exchange of photons but solely because of the fast change in boundary conditions. The new quantum phenomena like the dynamical Casimir effect and the dynamical Lamb effect open up new ways of controlling a qubit system and the entanglement within it.

The physical realization of a three qubits experiment in a non-stationary cavity involves superconducting qubits (as atoms), coupled to a coplanar waveguide (as the optical cavity), terminated by a superconducting quantum interference device (SQUID). In this way, one is able to tune the length of the cavity non-adiabatically just
 by modulating the SQUID's magnetic field, giving rise to the DLE. It would be physically impossible to realize such fast change in boundary conditions with real mirrors, for a system of atoms in a cavity. For this reason we need to use superconducting circuits.

Our interest in this kind of setup stems from the GHZ theorem \cite{ghz,ghz2}, a fundamental result
which states that for a system of three (or more) entangled objects the only framework suitable to make predictions is quantum mechanics.
It is important to mention that the experimental observation of the GHZ states
for three spatially separated photons was performed by measurements
of polarization correlations between three
photons~\cite{Zeilinger1999,Zeilinger2000}.

In this paper, a measure of the quantum entanglement of the system
by means of the conditional concurrence and the conditional residual tangle at the fixed amount of created photons is
introduced. The transition amplitudes and the
probabilities of excitation of the qubits due to the dynamical
Lamb effect are also calculated.
It was found that the conditional residual tangle is zero for any number of created photons except for the case of two created photons in the cavity. This means that after the change of boundary conditions the system can only end up in a state that shows pairwise entanglement of the qubits. The only exception happens for the case of exactly two created photons, where the system also shows simultaneous entanglement of all three qubits. For more than two created photons in the cavity there isn't much to say because such transitions are forbidden in our model.

It is important to remark that some of the quantities we have
obtained are increased when the frequency of the photons in the
final state $\omega_2$ approaches the transition frequency $E_0$ of
the qubits. This means that it is possible to increase the values of
those quantities if the frequency $\omega_2$ is tuned correctly by
modulating the SQUID's magnetic field in a coplanar waveguide. While the GHZ states were obtained experimentally due
to the quantum entanglement for  three spatially separated
photons~\cite{Zeilinger1999,Zeilinger2000}, the advantage of the
proposed setup with three superconducting qubits, coupled to a
coplanar waveguide is the possibility of tuning the GHZ states by
controlling the frequency of the cavity photons. Let us mention
that the choice of the parameters for our calculations is
constrained by the requirements imposed by perturbation theory.

We are mostly interested with the new quantity introduced here, the conditional residual tangle for each fixed number of created photons, because it tells us if the system of three qubits shows three-body entanglement, which is what we need to have a system that can be used to test the validity of quantum mechanics.

In fact, in Eq. (\ref{3state}) we found a final state of the system
showing a value of the conditional residual tangle different from
zero. So, for this final state the predictions of quantum mechanics
in the case of three entangled qubits can be tested. This is
important because it has been shown in Refs.~\cite{ghz2,ghz} that in this case there is no way to
replicate the predictions of quantum mechanics with a local
realistic theory. Therefore, to verify the validity of quantum
theory, we only need to verify that the experiments validate the
theoretical predictions we have made in this framework.

\acknowledgments
The authors are grateful to D. M. Greenberger, M. Hillery and W. Wootters for the valuable and stimulating discussions.

\appendix

\section{Eigenvalues and Eigenstates of the Hamiltonian}

\label{appendixa}

We are interested in corrections due to the DLE, which involves the parametric excitation of the qubit and the creation of cavity photons. Then, only $\hat{V}=\lambda
\sum_{j=1}^{3}\left( \hat{\sigma}_{j}^{+}a^{\dagger }+\hat{\sigma}%
_{j}^{-}a\right) $ is considered as a perturbation to the ground state Hamiltonian $\hat{H}_{0}$ in Eq. (\ref{ham022}). The qubit-photon interaction terms in $\hat{V}$ are assumed to be only a small perturbation to the non-interacting Hamiltonian $\hat{H_0}$ so that time-independent perturbation theory can be used to calculate the energy corrections (up to second order) and the wavefunction corrections (up to first order) of the eigenstates of the unperturbed Hamiltonian.

The background energy $E^{(0)}$ of the possible qubits states with $n$ photons in the cavity are:
\begin{eqnarray}
E_{n;0 0 0}^{(0)} &=& n\omega , \nonumber \\
E_{n;1 0 0}^{(0)} &=& E_{n;0 1 0}^{(0)} = E_{n;0 0 1}^{(0)} = n\omega + E_{0} , \nonumber \\
E_{n;1 1 0}^{(0)} &=& E_{n;1 0 1}^{(0)} = E_{n;0 1 1}^{(0)} = n\omega +2E_{0} , \nonumber \\
E_{n;1 1 1}^{(0)} &=& n\omega + 3E_{0} .\
\end{eqnarray}

Those are the energy eigenvalues of the unperturbed Hamiltonian $\hat{H}_{0}$, which describes a system of three qubits in a stationary cavity. They are the energy levels of the non-interacting system.

The eigenvalues of the Hamiltonian $\hat{H}$ can be calculated in the framework of time-independent second-order perturbation theory \cite{Landau}, using Eqs.~(\ref{ham022}) and~(\ref{ham223}). These will be the energy levels of the system when the boundary conditions are allowed to change non-adiabatically and qubit-photon interaction arises.

\begin{eqnarray}
\label{ener2} E_{n;0 0 0}^{(\lambda)} &=& n\omega+ 3\lambda^2\left(\frac{2E_{0}}{\omega^{2} - E_{0}^{2}}n - \frac{1}{\omega + E_{0} } \right), \nonumber \\
E_{n;1 0 0}^{(\lambda)} &=& E_{n;0 1 0}^{(\lambda)} = E_{n;0 1 0}^{(\lambda)} = n\omega + E_{0} +\lambda^{2}\left(\frac{2E_{0}} {\omega^{2} - E_{0}^{2}}n + \frac{E_{0} - 3\omega} {\omega^{2} - E_{0}^{2}}\right), \nonumber \\
E_{n;1 1 0}^{(\lambda)} &=& E_{n;1 0 1}^{(\lambda)} = E_{n;0 1 1}^{(\lambda)} = n\omega + 2E_{0} -\lambda^{2}\left(\frac{2E_{0}} {\omega^{2} - E_{0}^{2}}n + \frac{E_{0} + 3\omega} {\omega^{2} - E_{0}^{2}}\right), \nonumber \\
E_{n;1 1 1}^{(\lambda)} &=& n\omega + 3E_{0} -3\lambda^{2}\left(\frac{2E_{0}} {\omega^{2} - E_{0}^{2}}n + \frac{1} {\omega - E_{0}}\right). \
\end{eqnarray}
The first order correction is zero because the interaction term evaluated contains a lowering/raising operator of the qubits' excitation number and so the final state will always be a state orthogonal to the initial state (ground state). In the second order corrections, terms which do not depend on the number of photons $n$ can be seen as Lamb shift of the qubit's energy levels. In fact, the static Lamb effect does not change the number of photons since they are excited and then re-absorbed. Thus, the total Lamb shifts of the states of the three qubits system are:
\begin{eqnarray}
\label{Lamb} E_{L,0}^{(\lambda)} (\omega) \equiv E_{L,0 0 0}^{(\lambda)} (\omega) &=& -3\lambda^{2}\frac{1}{\omega + E_{0} }, \nonumber \\
E_{L,1}^{(\lambda)} (\omega) \equiv E_{L,1 0 0}^{(\lambda)}(\omega) &=& E_{L,0 1 0}^{(\lambda)} = E_{L,0 0 1}^{(\lambda)} =
\lambda^{2}\left(\frac{E_{0} - 3\omega} {\omega^{2} - E_{0}^{2}}\right), \nonumber \\
E_{L,2}^{(\lambda)} (\omega) \equiv E_{L,1 1 0}^{(\lambda)}(\omega) &=& E_{L,1 0 1}^{(\lambda)} = E_{L,0 1 1}^{(\lambda)} =   -\lambda^{2}\left(\frac{E_{0} + 3\omega} {\omega^{2} - E_{0}^{2}}\right), \nonumber \\
E_{L,3}^{(\lambda)} (\omega) \equiv E_{L,1 1 1}^{(\lambda)}(\omega) &=& -3\lambda^{2}\frac{1}{\omega - E_{0} }.
\end{eqnarray}
In this model, the total Lamb shift is the sum of the Lamb shifts of the individual qubits.
To get a sense of why the static Lamb shift is obtained as a second order correction to the energy levels, one can think of the Feynman's diagram for the qubit's self-energy: in the creation and absorption of a virtual photon two vertices (qubit-photon coupling) are involved. This means we are dealing with the second order of perturbation theory when we consider this phenomenon.

Using time-independent perturbation theory on Eqs.~(\ref{ham022}) and~(\ref{ham223}), one can get the eigenfunctions of the Hamiltonian $\hat{H}$. The corrections up to first order are:
\begin{eqnarray}
\label{st1} \left|n;0 0 0\right\rangle_{\lambda \omega} = \left|n;0 0
0\right\rangle + \frac{\lambda\sqrt{n}\left(\left|n-1;1 0
0\right\rangle + \left|n-1;0 1 0\right\rangle + \left|n-1;0 0 1\right\rangle\right)}{\omega - E_{0}} + \nonumber \\ - \frac{\lambda\sqrt{n+1}\left(\left|n+1;1 0 0\right\rangle +
\left|n+1;0 1 0 \right\rangle + \left|n+1;0 0 1 \right\rangle\right)}{\omega + E_{0}}, \nonumber \\ \nonumber \\
\left|n;1 0 0\right\rangle_{\lambda \omega} = \left|n;1 0
0\right\rangle + \frac{\lambda\left(\sqrt{n}\left|n-1;1 0
1\right\rangle + \sqrt{n}\left|n-1;1 1 0\right\rangle - \sqrt{n+1} \left|n+1;0 0 0\right\rangle\right)}{\omega - E_{0}}
+ \nonumber \\ - \frac{\lambda\left(\sqrt{n+1}\left|n+1;1 0 1\right\rangle +
\sqrt{n+1}\left|n+1;1 1 0 \right\rangle -\sqrt{n} \left|n-1;0 0 0 \right\rangle\right)}{\omega + E_{0}}, \nonumber \\ \nonumber \\
\left|n;0 1 0\right\rangle_{\lambda \omega} = \left|n;0 1 0\right\rangle + \frac{\lambda\left(\sqrt{n}\left|n-1;1 0
1\right\rangle + \sqrt{n}\left|n-1;1 1 0\right\rangle - \sqrt{n+1} \left|n+1;0 0 0\right\rangle\right)}{\omega - E_{0}} + \nonumber \\ - \frac{\lambda\left(\sqrt{n+1}\left|n+1;1 0 1\right\rangle +
\sqrt{n+1}\left|n+1;1 1 0 \right\rangle -\sqrt{n} \left|n-1;0 0 0 \right\rangle\right)}{\omega + E_{0}}, \nonumber \\ \nonumber \\
\left|n;0 0 1\right\rangle_{\lambda \omega} = \left|n;0 0
1\right\rangle + \frac{\lambda\left(\sqrt{n}\left|n-1;1 0
1\right\rangle + \sqrt{n}\left|n-1;1 1 0\right\rangle - \sqrt{n+1} \left|n+1;0 0 0\right\rangle\right)}{\omega - E_{0}}
+ \nonumber \\ - \frac{\lambda\left(\sqrt{n+1}\left|n+1;1 0 1\right\rangle +
\sqrt{n+1}\left|n+1;1 1 0 \right\rangle -\sqrt{n} \left|n-1;0 0 0 \right\rangle\right)}{\omega + E_{0}}, \nonumber \\ \nonumber \\
\left|n;1 1 0\right\rangle_{\lambda \omega} = \left|n;1 1
0\right\rangle + \frac{\lambda\left(\sqrt{n}\left|n-1;1 1
1\right\rangle - \sqrt{n+1}\left|n+1;1 0 0\right\rangle - \sqrt{n+1} \left|n+1;0 1 0\right\rangle\right)}{\omega - E_{0}}
+ \nonumber \\ + \frac{\lambda\left(\sqrt{n}\left|n-1;1 0 0\right\rangle +
\sqrt{n}\left|n-1;0 1 0 \right\rangle -\sqrt{n+1} \left|n+1;1 1 1 \right\rangle\right)}{\omega + E_{0}}, \nonumber \\ \nonumber \\
\left|n;1 0 1\right\rangle_{\lambda \omega} = \left|n;1 0
1\right\rangle + \frac{\lambda\left(\sqrt{n}\left|n-1;1 1
1\right\rangle - \sqrt{n+1}\left|n+1;1 0 0\right\rangle - \sqrt{n+1} \left|n+1;0 1 0\right\rangle\right)}{\omega - E_{0}}
+ \nonumber \\ + \frac{\lambda\left(\sqrt{n}\left|n-1;1 0 0\right\rangle +
\sqrt{n}\left|n-1;0 1 0 \right\rangle -\sqrt{n+1} \left|n+1;1 1 1 \right\rangle\right)}{\omega + E_{0}}, \nonumber \\ \nonumber \\
\left|n;0 1 1\right\rangle_{\lambda \omega} = \left|n;0 1
1\right\rangle + \frac{\lambda\left(\sqrt{n}\left|n-1;1 1
1\right\rangle - \sqrt{n+1}\left|n+1;1 0 0\right\rangle - \sqrt{n+1} \left|n+1;0 1 0\right\rangle\right)}{\omega - E_{0}}
+ \nonumber \\ + \frac{\lambda\left(\sqrt{n}\left|n-1;1 0 0\right\rangle +
\sqrt{n}\left|n-1;0 1 0 \right\rangle -\sqrt{n+1} \left|n+1;1 1 1 \right\rangle\right)}{\omega + E_{0}}, \nonumber \\ \nonumber \\
\left|n;1 1 1\right\rangle_{\lambda \omega} = \left|n;1 1
1\right\rangle + \frac{\lambda\sqrt{n}\left(\left|n-1;1 1
0\right\rangle + \left|n-1;1 0 1\right\rangle + \left|n-1;0 1 1\right\rangle\right)}{\omega - E_{0}}
+ \nonumber \\ - \frac{\lambda\sqrt{n+1}\left(\left|n+1;1 1 0\right\rangle +
\left|n+1;1 0 1 \right\rangle + \left|n+1;0 1 1 \right\rangle\right)}{\omega + E_{0}},
\end{eqnarray}
where $n$ is the number of photons.
We note that each state calculated in this way is a combination of the state itself and all other states that differ from it by one qubit excitation number. This is due to the perturbation we have considered in making our calculations, whose terms are proportional only to a single qubit's excitation raising/lowering operator ($\sigma_i^{+/-}$).

\end{document}